# Modeling Spatial Equilibrium in Cities: the Isobenefit Lines

Luca D'Acci
lucadacci@gmail.com

**Abstract**
I propose and briefly define the concept of *Urban Isobenefit Lines* by using functions as easy as efficient, whose results can offer a rich tool to use into spatial equilibrium analysis involving cities. They are line joining urban points with equal level of positional advantage from city amenities. The results which one obtain by implementing a chosen function, gave specific scenarios: numerically described by indicators and graphically visualized by efficient city matrix views. This is also a theoretical concept for the Urban Economics theory and Spatial Equilibrium analysis in cities.

**Keywords**: Modeling Spatial Equilibrium, Urban Economy, City Amenities, Urban Centralities, Property Value, Urban Quality of Life.

## 1. Spatial Equilibrium in Cities

The basis of economic approaches of cities is the assumption that locations are chosen, and that those choices, although not completely rational, are not entirely irrational too [8].
Citizens are not randomly spread in the city areas: they chose them. When they select a particular location for their residences, their decisions are a balanced comparative sum of the following factors: location of work, family and friends, housing cost, quality of the area, distance from the city centre and from other city centralities, public services, public goods, urban amenities.
This balance process is in part conscious in part not, and in some models it is considered identical for everyone, while some others [5] venture to formulize the heterogeneity of the citizens' minds. "Individual choice over locations produces the single most important concept in urban or regional economics: the spatial equilibrium" [8]. The background of this statement comes from the supposition that if analogous citizens chose to live in *n* different city areas, then each of these *n* urban points are offering a comparable sum of advantages and disadvantages.
We can write this general idea of spatial equilibrium in the following simplified equation:

$$U(i, o, a) = k \tag{1}$$

Where citizens maximize a utility function (*U*) against their financial budget (*i* = input money: wage) and their costs (*o* = output money: housing costs, commuting costs, taxes ...) in relation with the advantages (*a*) which they receive by living in a specific urban point.
When one find that/those urban point/s which allow him that right level of utility (*k*) he does not have the necessity to change his residence to another urban point.
"a" is a vector containing the advantages (as well as disadvantages) offered from the urban points, such as distance from family and friends residence and from job, and amenities such as social context, pollution, noise, parks, shops, pedestrian areas, aesthetics and quality of streets and squares, crime, comfort of public transport, cycling paths, schools, etc.
Inside this spatial equilibrium pattern, I investigate that part of the citizen location decision (a), connected to the advantage coming from the amenities of the city areas.
This study is limited on Equilibrium while Evolution must be studied by appropriate methods such as Statistical Physics and the Games Theory [1,2].

## 2. Isobenefit Lines

Translating in mathematical terms the elementary concept shown above, I obtain a city map with isolines of the spatial advantage coming from urban amenities. Those isolines (or contours) join the points of the city with a same positional advantage connected to the city amenities, therefore, their shape is strongly affected from the quantity, quality and spatial distribution of the amenities, and, especially, from the 'cost' of moving, that I count as a pondered sum of three variables: cost, comfort and speed of moving.

To first step to get the isolines is to separate *Punctual Benefit* (*A*) of an Amenity, and its *Distributed Benefit* (*B*), [3,4]: *A* concerns the benefit obtained from the citizen by using the amenity (park, square, etc); *B* refers to the benefit of every urban points given from *A*. Thanks to methods shown from the literature we can attempt to quantify $A^1$ [6], while *B* is associated to the citizens' facility of using the city's attractions and depends on the benefit (*A*) offered by the attraction (*i*), the distance (*d*) between the citizen location (*k*) and the attraction (*i*), and the cost/comfort/speed – we can call these three factors *E* (Efficiency of moving) – of reaching the attraction:

$$B_{i,k} = (A_i \cdot E) \cdot (d_{i-k} + E)^{-1}$$

(2)

Where $B_{i,k}$ is the Benefit in a *k* urban point, generated by an *i* attraction having an *A* level of Punctual Benefit, and $d_{i-k}$ is their distance. In order to have logical[2] results *E* is suggested to not be high. *E* will vary throughout the city in relation with the public transport structure (underground, traffic of street, cycle paths, etc.). By inverting the way we read *E*, we can also chose other functions like:

$$B_{i,k} = A_i \cdot e^{-E \cdot (d_{i-k})^2}$$

(3)

$$\text{or } B_{i,k} = A_i \cdot e^{-E \cdot (d_{i-k})}$$

(4)

or even give different spatial distributional effects (that means chose different function), for each kind of amenities.

Independently from the function we prefer to use, I define *Curves of Positional Isobenefit* the lines that join the urban points (*k*) with the same level of *B*. We can also call them *Isobenefit Contours*, or *Isobenefit Lines*.

---

[1] We should numerically judge how much an attraction can satisfy its own definition and pretension to be an 'Attraction', and not for a specific citizen, but for the majority of them. For this reason it could be reasonable to judge *A* by referring to the usual, average number of citizens (not tourists) using the attraction and by then comparing each amenity with the best place/s in the city and with the neutral ones. In a similar way, but in different contest and aim, urban economists have often been interested in using population levels as a measure of urban success. High levels of population "tell us that people are voting with their feet to move to a particular place" (Glaeser 2008). Certainly there is no doubt about the relativity, and then the validity, of our own preferences also if divergent from other people, or even from the average peoples preferences: that which for a person can be a wonderful attraction, i.e. a shopping mall, for another can be a boring, consumerist place. Idem for the judgment of amenities such as parks, historical areas, and so forth.

[2] *E* should not be chosen too high, in order to assure a shape of the function where the gradient of the benefit, against the distance, varies considerably (graphically speaking, closer to a parabolic, hyperbolic shape rather than to a linear one). Otherwise, following the equation, a point in the middle of two amenities could even result with a similar benefit than a point in front of one of the two amenities.

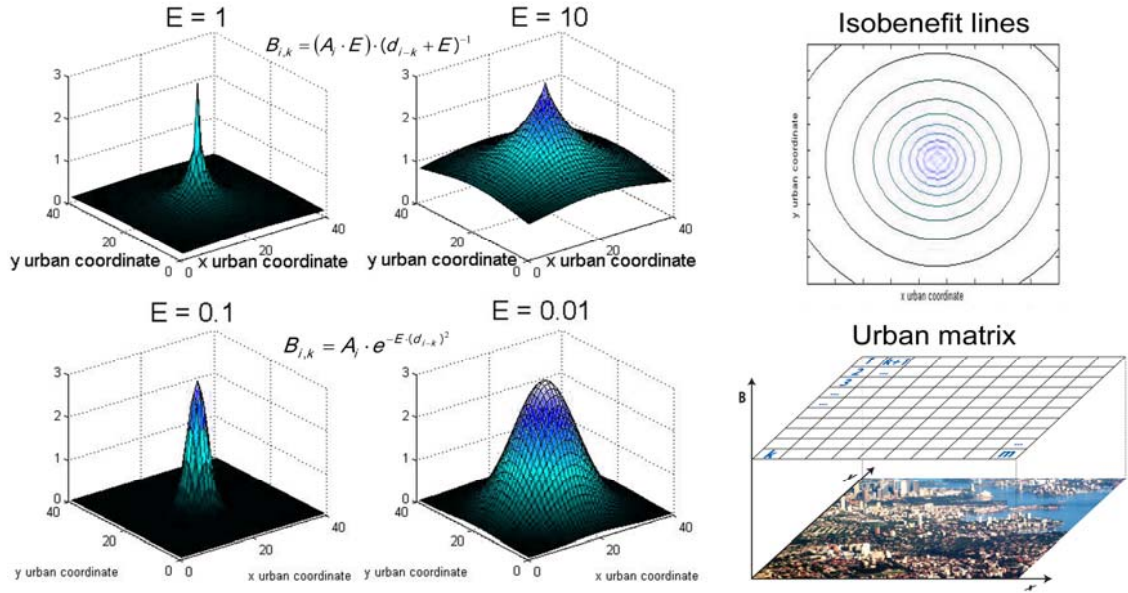

Fig. 1. Isobenefit lines.

If we want to calculate the benefit by taking into account the number of beneficiaries, we just multiply equation 2 by $D_k$ (population density in $k$).

## 3. Isobenefit Orography

Considering all the attractions present in a city, $B_k$ is the benefit of the $k$ urban point given by all the $n$ attractions in the city:

$$B_k = \sum_{i=1}^{n} B_{i,k} = \sum_{i=1}^{n} \frac{A_i \cdot E}{d_{i-k} + E}$$

(5)

We can have different isobenefit scenarios depending on the planimetric distribution of the amenities throughout the city, and on their attractiveness level. I define the 3D visualization of the isobenefit lines as *Isobenefit Orography* (Fig. 2 on the left).

Calling $M$ the matrix of $m$ elements (cells whereby we have divided the city), where the generic $k$ element has a value $B_k$, the $m_{xy}$ element (the $k$ urban cell[3]) of $M$ is [3]:

$$m_{xy} = B_k = \sum_{i=1}^{n} (A_i \cdot E)[((x_k - x_i)^2 + (y_k - y_i)^2)^{1/2} + E]^{-1}$$

(6)

The total Benefit is given by:

$$B_t = \sum_{i=1}^{n}\sum_{k=1}^{m} B_{i,k} = \sum_{i=1}^{n}\sum_{k=1}^{m} A_i \cdot E / (d_{i-k} + E)$$

(7)

Or, in differential maths:

$$B_t = \sum_{i=1}^{n} \int_{a_i}^{b_i}\int_{c_i}^{d_i} \{(A_i \cdot E)[((x - x_i)^2 + (y - y_i)^2)^{1/2} + E]^{-1}\} dx dy$$

$$a_i = x_{min} - x_i \quad b_i = x_{max} - x_i$$
$$c_i = y_{min} - y_i \quad d_i = y_{max} - y_i$$

(8)

in which $x_{min}$, $x_{max}$ and $y_{min}$, $y_{max}$ are the minimum and maximum coordinates of the urban space, in this case approximated as a rectangle, and $x_i$ and $y_i$ are the coordinates of the $i$ attractions.

---
[3] I will intend as benefit of the cell the benefit in its geometrical centre.

The *Uniformity Coefficient* (*U*), is given by:

$$U = 1 - \sqrt{\frac{\sum_{k=1}^{m}\left(B_k - \frac{1}{m}\sum_{k=1}^{m}B_k\right)^2}{m}} \bigg/ \frac{1}{m}\sum_{k=1}^{m}B_k$$

(9)

*U* is a number less or equal to 1 (maximum uniform distribution). It is relative, not absolute, therefore can also be used to compare different cities among each other or a same city before and after some transformations in its scenario[4].

In Fig. 2, citizens living in points 1 and 7 are in front of an amenity (A4 and A2); citizens in points 5 and 3, are a bit more distant from their closest amenity (A3 and A1), but the amenity has an higher attractiveness, that means that the disadvantage to live not exactly in front of the amenity is compensated from the higher benefit (A, Punctual Benefit) that the citizen enjoys when reaching the amenity. In the same way, citizens living in points 2, 4, 6 and 8 are more distant from their closest amenity, but enjoy the advantage to reach not one, but two or three amenities with the same effort (time-cost of commuting).

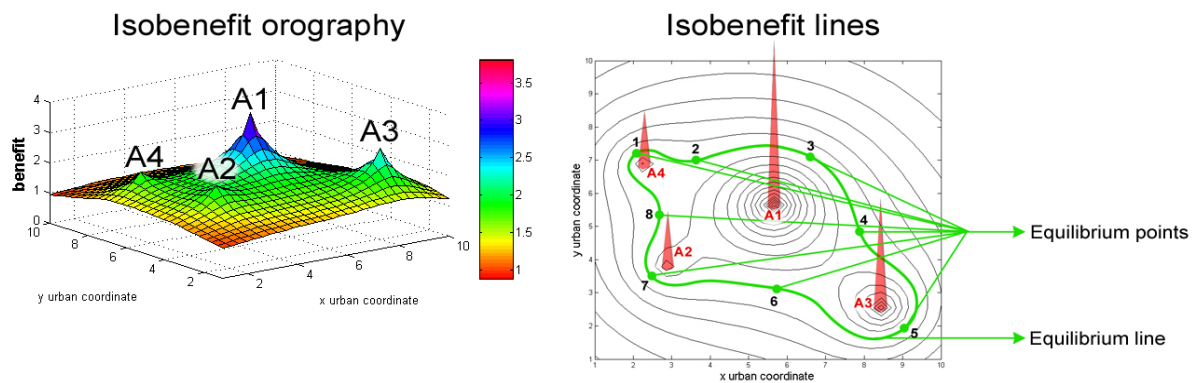

Fig. 2. Isobenefit Lines as Spatial Equilibrium Lines.

We can also build *Personal Isobenefit Lines* connected to a specific citizen. In fact, Isobenefit Lines, can deeply vary among people, and even among different state-age of a same person. For example, looking at Fig. 2, children and elderly will prefer living in point 1 rather than points 2, 3, 4, 5, 6 and 8. On the contrary, others could prefer living in 2, 3, 4, 5, 6 and 8 rather than in point 1 or 7, because they could prefer the variety offered from the availability to access more than one amenity (and with a greater level of attractiveness), even paying the cost of not enjoying living very close to any amenity (although with a low attractiveness). If we build Personal Isobenefit Lines, *E* can also be read as the personal propensity to move, or the personal preference for the *Variety advantage* (*Va*) (directly proportional to *E* in eq. 5) rather than for the *Proximity advantage* (*Pa*) (inversely proportional to *E*); where *Va* is the advantage to be able to reach numerous amenities, and *Pa* the advantage to be very close to one amenity.

---

[4] *U* measures the uniformity of the *positional* advantage of each urban point. It does not quantify the uniformity of the amenities location throughout the city. When we compare different scenarios, *U* should be read not alone, but together with other indicators such as the total, medium, maximum and minimum value of *B*. We imagine a radiocentric city with all the best attractions uniformly distributed on the external circular crown (scenario 1), or with its attractions uniformly distributed throughout all the city area (scenario 2). It could (depending on *E*) even result *U*1>*U*2, but reading the results better, we would notice that the level of *B* is higher in each point in scenario 2 rather than in 1. *U* could result higher in 1 than in 2 because the central part has always a positional advantage due to its geometrical position, even if it does not have any amenity in front of it (scenario 1). In scenario 1, it has an advantage of "variety"; it can get all the amenities with the same effort (distance). If we put amenities uniformly covering all the city area (scenario 2), the positional advantage of the centre becomes even higher in comparison with the periphery (*U*1>*U*2). In equation 5, the higher *E*, the higher this distortion of the lecture of *U* (and *B*) could be. For this it is suggested a range of low values of *E*. The higher *E* the more the equation 'weighs' the 'variety' to enjoy numerous amenities, rather than the advantage of the proximity of one amenity. If we want to consider also the disamenities (noisy, dirty streets, old factory in the city, cemetery, etc.) we can add a negative value of A, but in this case *U* will give some problems. Therefore, for *U*, we should separately consider amenities and disamenities.

Fig. 3 (second row) shows *B* before and after the recent urban transformations of Turin [4]: the abscissa is $B_k$, the ordinate is the frequency. By transforming *M* into a vector, we can plot $B_k$ (ordinate) of each urban cell (abscissa) (Fig. 2, first row).

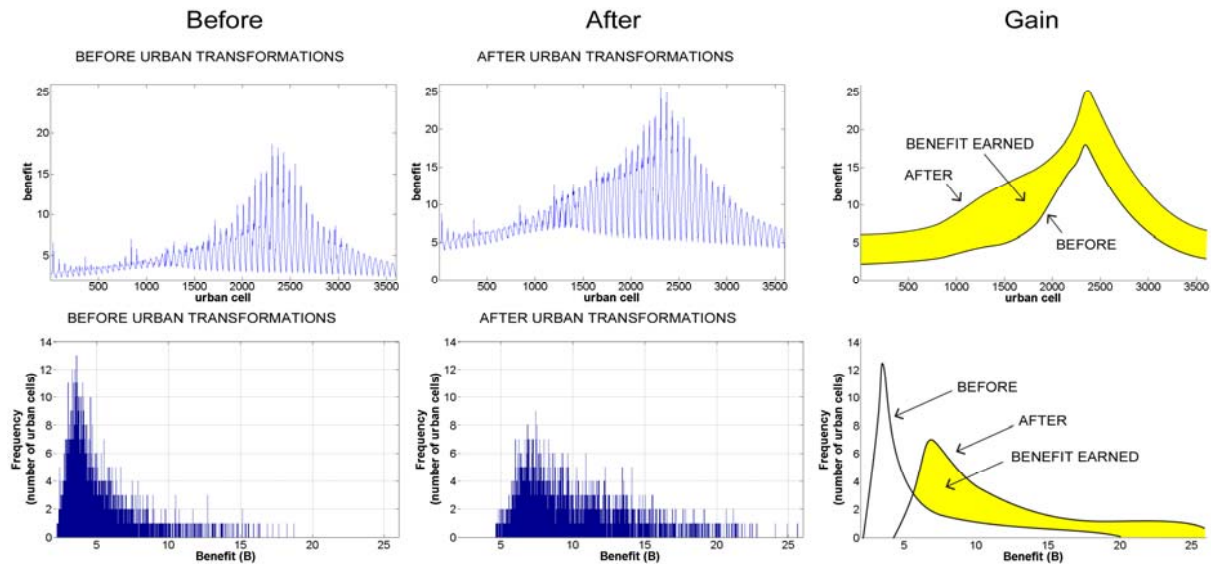

Fig. 3 Gain of Benefit.

**4. Conclusion**

The classical urban economic theory – Malthus, von Thünen, Christaller, Losh, Alonso, Muth, Mills – is based on the trade-off between accessibility and space: urban points which have better access to the Central Business District (CBD) have higher land values. Usually, in the standard case considered in these original theories, urban amenities fall with the distance from the central business district (CBD). This tradition, which defines the theoretical nucleus of the urban economy, has continued to grow through the use of more complex models [7,9]. However, nowadays we assist a general aim and tendency to generate multicentre cities where the spatial variation in amenities allocation and centralities throughout the city, creates a richer pattern. Therefore, the above mentioned phenomena will not be described by a monotonic function of the distance to the CBD, but by a more elaborated function showed in this study. There is a strong correlation between Isobenefit Lines and Property value [4]. I also presented an indicator to assess the homogeneity of the amenities spatial distribution benefit.